\documentclass[useAMS,usenatbib]{mn2e}
\newcommand{\etal}{\mbox{et\ al.\ }}

\newcommand{\plusminus}[2]{\ensuremath{^{+#1}_{-#2}}}
\usepackage{rotating}

\title[Constraining the energy budget of GRB\,080721]
{Constraining the energy budget of GRB\,080721}

\author[Starling, Rol, van der Horst \etal 2009]
{R.L.C. Starling$^1$\thanks{e-mail: rlcs1@star.le.ac.uk}, E. Rol$^{1,2}$, A.J. van der Horst$^3$, S.-C. Yoon$^4$, V. Pal'shin$^5$, \and 
C. Ledoux$^6$, K.L. Page$^1$, J.P.U. Fynbo$^7$, K. Wiersema$^1$, N.R. Tanvir$^1$, P. Jakobsson$^8$, \and C. Guidorzi$^{9,10}$,
 P.A. Curran$^{11}$,  A.J. Levan$^{12}$, P.T. O'Brien$^1$, J.P. Osborne$^1$, \and D. Svinkin$^5$,
 A. de Ugarte Postigo$^{6}$, T. Oosting$^{13}$ and I.D. Howarth$^{14}$
\\$^1$Department of Physics and Astronomy, University of Leicester, University Road, Leicester LE1 7RH, UK.\\$^2$Astronomical Institute, University of Amsterdam, PO Box 94249, 1090 GE, Amsterdam, The Netherlands.\\$^3$NASA Postdoctoral Program Fellow, NSSTC, 320 Sparkman Drive, Huntsville, AL 35805, USA.\\$^4$UCO/Lick Observatory, Department of Astronomy and Astrophysics, University of California, Santa Cruz, CA 95064, USA.\\$^5$Ioffe Physico-Technical Institute, Laboratory for Experimental Astrophysics, 26 Polytekhnicheskaya, St Petersburg 194021, Russian Federation.\\$^6$European Southern Observatory, Alonso de C\'ordova 3107, Casilla 19001, Vitacura, Santiago, Chile.\\$^7$Dark Cosmology Centre, Niels Bohr Institute, University of Copenhagen, Juliane Maries Vej 30, DK-2100 Copenhagen \O, Denmark.\\$^8$Centre for Astrophysics and Cosmology, Science Institute, University of Iceland, Dunhagi 5, 107 Reykjav\'ik, Iceland.\\$^9$INAF$-$Osservatorio Astronomico di Brera, Via E. Bianchi 46, I-23807 Merate (LC), Italy.\\$^{10}$Physics Department, University of Ferrara, via Saragat 1, 44100, Ferrara, Italy.\\$^{11}$Mullard Space Science Laboratory, University College London, Holmbury St. Mary, Dorking, Surrey RH5 6NT, UK.\\$^{12}$Department of Physics, University of Warwick, Coventry CV4 7AL, UK.\\$^{13}$Department of Astrophysics, University of Nijmegen, PO Box 9010, 6500 GL, Nijmegen, The Netherlands.\\$^{14}$Department of Physics and Astronomy, University College London, Gower Street, London WC1E 6BT, UK.}

\begin{document}
\date{Accepted . Received ; in original form }
%\bigskip

\pagerange{\pageref{firstpage}--\pageref{lastpage}} \pubyear{}

\maketitle

\label{firstpage}

%%***************************************************************************
%%  ABSTRACT
%%***************************************************************************

\begin{abstract}
We follow the bright, highly energetic afterglow of {\it Swift}-discovered GRB\,080721 at $z=2.591$ out to 36 days or 3$\times$10$^{6}$~s since the
trigger in the optical and X-ray bands. We do not detect a break in the late-time light
curve inferring a limit on the opening angle of $\theta_j \ge 7.3^{\circ}$ and setting tight constraints on the total energy
budget of the burst of $E_{\gamma} \ge 9.9\times 10 ^{51}$ erg within the fireball model. To obey the fireball model closure relations the GRB jet must be expanding into a homogeneous surrounding medium and likely lies behind a significant column of dust.
The energy constraint we derive can be used as observational input for models of the progenitors of long gamma-ray bursts: we discuss how such high collimation-corrected energies could be accommodated with certain parameters of the standard massive star
core-collapse models. We can, however, most probably rule out a magnetar progenitor for this GRB which would require 100\% efficiency to reach the observed total energy. 
\end{abstract}

\begin{keywords}
gamma-rays:bursts
\end{keywords}

\section{Introduction}
The discovery \citep{VanParadijs,Costa} of afterglows to long-duration gamma-ray
bursts (GRBs) showed them to occur in star-forming galaxies at high
redshifts. Many aspects of the observed GRB behaviour could be
explained reasonably well by the relativistic fireball models, in
which the prompt emission is largely produced by shocks internal to
the outflow, and the long-lived afterglow by the outflow impacting and
shocking the external ambient medium \citep{MeszarosRees,Sari}. However, it was early appreciated
that if they emitted isotropically then the radiative energies implied
for some bursts would be very large (in excess of a Solar rest mass in
the case of GRB\,990123, for example; \citealt{Kulkarni}). Since no known mechanism can produce high-energy photons with
efficiency approaching 100\%, the total explosive energy required
would be even greater, implausibly large for a stellar core-collapse
powered event. 

The energetics argument led to the expectation that the GRB outflow must
be confined to a jet, reducing the overall energy requirements.  In
that case a sufficiently massive core collapsing to a black-hole might
produce an outflow which could pierce a hydrogen-stripped envelope to
still produce a relativistic jet \citep{MacFadyen}.
This picture received support from the observation of concurrent
supernova events at GRB sites for some (low-redshift) low-luminosity
bursts \citep{Galama,Hjorth,Stanek,Pian}. The class of GRB-supernovae could be explained by
the explosion of a Wolf-Rayet star, and it has been widely assumed that this
progenitor must account for most if not all long-GRBs.

A strong prediction of all collimated models is that the observed
light curve should exhibit an achromatic break, visible in both optical and X-ray light curves, when the relativistic
outflow slows to the point that the Doppler beaming angle becomes
wider than the opening angle of the jet \citep{Rhoads}: the
later the break time, the wider the jet. In the pre-{\it
Swift} era a number of bursts with good afterglow observations showed
breaks in the optical, at times ranging from a few hours to a few
days. Furthermore, it was contended
that the implied jet opening angles anti-correlated with the
isotropic-equivalent $\gamma$-ray energies $E_{\rm iso}$ \citep{Frail} -- in other words, corrected for
collimation nearly all bursts seemed to be accessing a rather standard
energy reservoir (to within about an order of magnitude), potentially usable as standard
candles at high redshift. 
While the total rotational energies available from a collapsing core may be of
order 10$^{54}$ erg, the maximal extraction of energy is unlikely to exceed
10$^{51}$--10$^{52}$ erg \citep{Paczynski,Popham}. With this in mind, the standard energy reservoir with
energies of $\sim$10$^{51}$ erg appeared very attractive.

Unfortunately the X-ray data available pre-{\it Swift} only provided
definitive evidence of (single) breaks for a few bursts. One of the
expectations of {\it Swift} was that many more bursts would be
seen with clear breaks in the X-ray coincident with optical breaks.
After more than four years of operations this is not what has been found.
In fact, the situation is much more confusing, with X-ray light curves
exhibiting complex behaviour including flares and plateaux \citep{Nousek,OBrien,Willingale}. When
late breaks (beyond a few hours) are seen they rarely coincide in time
or degree with optical breaks, and are hard to reconcile with simple
fireball model predictions \citep[e.g.][]{Panaitescu06}, though there are notable exceptions. It is possible that from X-ray data alone a number of jet breaks, particularly those at late times or those which turn over rather smoothly, may go undetected through misinterpretation of their light curves. These so-called `hidden' breaks may be revealed on comparison of X-ray light curves with optical light curves \citep{Curran}, but the emphasis remains on X-ray light curves for break detection owing to the large numbers of
well-sampled X-ray light curves now available. 

Most crucially, in many GRBs no temporal break is seen at all to late
times, implying much more energy than
expected. The most extreme example to-date is that of GRB\,060729,
which continued a smooth X-ray power-law decline for 125 d,
implying an opening angle greater than 28 degrees \citep{Grupe}. 
This source continues to be monitored in the X-rays with {\it Chandra} and was still detected one year after the GRB event (Grupe et al. 2009). The light curve now shows a possible break at $\sim$290 d: the nature of this break is not yet known. 
While the modest luminosity and low redshift of GRB\,060729 ($E_{\rm iso}\sim 7 \times 10^{51}$ erg, \citealt{Grupe}) make it unsuitable to test collapsar models to their limits,
such observations provide a crucial guide to further theoretical
developments. If such high collimation-corrected energies continue to be inferred, then
the situation looks bleak for simple core-collapse models as the
explanation for the high-luminosity long-GRBs. In fact, it would
suggest a different progenitor is required for high- and
low-luminosity long-bursts. On the other hand, if late breaks are seen, although the jet model is not proven this could provide evidence
for an upper envelope to the energy tapped by GRBs which may still be consistent with that available in principle from
core-collapse, but would require a new understanding of how such
high efficiencies of radiative emission could be achieved.

In this paper we follow a high redshift GRB with a high isotropic energy out to late times to search for a jet break and thereby measure the total energy budget of the prompt $\gamma$-ray emission. In Section 2 we describe the $\gamma$-ray, X-ray and optical observations and analyses. We derive the GRB fluence in Section 3. In Section 4 we characterise the afterglow spectra and light curves and compare the results to the fireball model in Section 5 discussing the possibility of an early jet break. In Section 6 we calculate limits on the jet opening angle and collimation-corrected energy of this GRB. In Section 7 we note the caveats associated with our results, discuss the energetics of this GRB in relation to current popular progenitor models and compare this GRB to other bursts of interest. In Section 8 we conclude. 

\section{Observations}
Throughout this paper we use the conventions $F \propto \nu^{-\beta} t^{-\alpha}$ for temporal and spectral power law models. Errors are quoted at the 90\% confidence level unless otherwise stated.
\begin{figure}
\begin{center}
\includegraphics[width=8cm, angle=0]{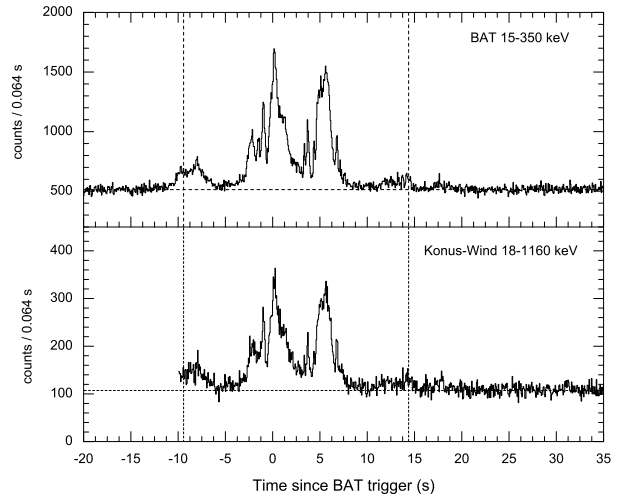}
\caption{
{\it Swift} BAT (top panel) and {\it Konus-Wind} (lower panel) 64~ms light curves of the prompt emission. The dashed vertical lines indicate the time interval adopted for joint spectral fits. }
\label{batkonuslc}
\end{center}
\end{figure}

\subsection{$\gamma$-ray}
The {\it Swift} satellite triggered on GRB\,080721 on 2008 July 21 at 10:25:16.979 UT (T$_{\rm 0,BAT}$). The Burst Alert Telescope (BAT) $\gamma$-ray light curve showed approximately six peaks, 3 strong followed by 3 weak, spanning T$_{\rm 0,BAT}$-11~s to T$_{\rm 0,BAT}$+370~s and the burst duration as defined by the T$_{90}$ parameter was 16.2$\pm$4.5~s \citep{Marshall}. The BAT data were reduced using the {\it Swift} analysis software version 29 (released 2008 June 29, part of HEAsoft 6.5.1).

The {\it Konus-Wind} experiment triggered on the
prompt emission of GRB\,080721 beginning at 10:25:10.927 UT (T$_{\rm 0,KW}$).
The propagation delay from {\it Swift} to {\it Konus-Wind} is 3.375~s for this GRB, i.e.,
correcting for this factor one sees that the {\it Konus-Wind} trigger time corresponds
to T$_{\rm 0,BAT}$-9.427~s. From T$_{\rm 0,KW}$ to T$_{\rm 0,KW}$+482.560~s, 64 spectra in 101 channels
(from ~20 keV to ~15 MeV) were accumulated on time scales varying from 64~ms
near the trigger time to 8.192~s by the time the signal became undetectable \citep{Golenetskii}. Data were processed using standard {\it Konus-Wind} analysis tools. The 20~keV--7~MeV peak flux is ($2.11\pm0.35$)$\times$10$^{-5}$ erg cm$^{-2}$ s$^{-1}$ (using the spectrum accumulated over
the main peak, from T$_{\rm 0,KW}$+8.448~s to T$_{\rm 0,KW}$+10.494s). The corresponding isotropic peak luminosity is $L_{\rm iso,max} = (1.11 \pm 0.18)\times 10^{54}$ erg s$^{-1}$.

The $\gamma$-ray light curves from both BAT and {\it Konus-Wind} are shown in Fig. \ref{batkonuslc}.

\subsection{X-ray}
The {\it Swift} X-Ray Telescope (XRT) slewed to the burst and began observations in Windowed Timing (WT) mode 113~s after the BAT trigger, detecting the tail of the prompt emission and the X-ray afterglow, and transitioned to Photon Counting (PC) mode in the second orbit of data collection. With XRT we have monitored the bright afterglow out to 1.4$\times$10$^6$~s or 16~d \citep{Marshall}.
The XRT data were reduced using the {\it Swift} analysis software version 29, as described in \cite{Evans,Evans2009}. Light curves use dynamic binning with a minimum binsize of 5~s and spectra are grouped such that a minimum of 20 counts lie in each bin.

We triggered our target-of-opportunity programme on the {\it XMM-Newton} satellite to obtain continued observations of GRB\,080721 once the X-ray afterglow flux had fallen below the sensitivity limits of the {\it Swift} XRT.  
The afterglow is clearly detected with the European Photon Imaging Cameras' (EPIC) PN in a 12~ks observation performed on 2008 August 12 and again in a 73~ks observation beginning 2008-08-26 and ending 2008-08-27.
The {\it XMM-Newton} data were reduced using the standard {\it XMM-Newton} Science Analysis Software version 20080701-1801. The second observation is partially affected by high background so we removed periods of background with $>$~1~count~s$^{-1}$ in the PN data leaving 46.5~ks of useful exposure time.
We used the tool {\small EDETECT-CHAIN} to identify the source position and measure a background-subtracted count rate. We then used {\small ESPECGET} to create spectra, and obtain the flux at each epoch by assuming a spectral shape identical to that of the best fit to the time-averaged {\it Swift} XRT PC mode spectrum. 
\subsection{Optical}
The UltraViolet-Optical Telescope (UVOT) on-board {\it Swift} began observations 118~s after the BAT trigger and found a bright optical afterglow in all optical filters, while the source was not detected in the UV filters \citep{Marshall}. Here we will focus on the $white$ and $v$ bands where temporal coverage was most comprehensive. 

We initiated ground-based observations once GRB\,080721 became visible from the Roque de Los Muchachos Observatory, La Palma. Observations were taken using the Nordic Optical Telescope (NOT) and the Liverpool Telescope (LT), followed several hours later by observations from Paranal Observatory with the Very Large Telescope (VLT) UT2 (Kueyen). After the initial observations, we initiated our programme for late-time light curve monitoring, obtaining further observations with the LT, the Isaac Newton Telescope (INT), the William Herschel Telescope (WHT) and the VLT UT1 (Antu). All optical afterglow photometry is listed in Table \ref{optdata}. Ground-based optical photometry have been reduced in the standard fashion using IRAF \citep{Tody}. For the photometric calibration, we used the zeropoints for the VLT FORS2, provided by the observatory\footnote{\texttt{http://www.eso.org/observing/dfo/quality/FORS2/qc/photcoeff\linebreak /photcoeffs\_fors2.html}}, after verifiying the corresponding nights were photometric using standard stars. For the $r$ and $i$ filters, we converted the $R$ and $I$ zeropoints to SDSS filters using the conversion at the SDSS webpages\footnote{\texttt{http://www.sdss.org/dr6/algorithms/sdssUBVRITransform.html\linebreak \#Lupton2005}}. We adopt a Galactic extinction of $E(B-V) = 0.102$ mag \citep{Schlegel}.

A low resolution spectrum was taken at the NOT from which we measured a redshift of $z=2.591 \pm 0.001$ reported in \cite{Jakobsson}, confirming an earlier redshift determination from independent data \citep{DAvanzo}. Starting on 2008 July 22.121 (0.7~d post-burst), we obtained a
sequence of higher resolution spectra using the VLT FOcal Reducer and low dispersion Spectrograph (FORS) 1 in long-slit spectroscopy
mode with a 1\farcs0 wide slit. The progression of grisms used was
300V (1200~s), 1200B (1800~s) and 600V+GG435 (1800~s). The
individual spectra were cosmic ray cleaned using the method of \cite{vanDokkum}. The seeing stayed relatively stable during the observations, between 1\farcs1 and 1\farcs3, yielding a
spectral resolution of 13.4 (300V), 2.9 (1200B) and 5.9 (600V+GG435)
\AA\ full width at half maximum (FWHM). Flux calibration was performed using an observation of the
standard star BPM16274. From the optical spectrum we derive a neutral hydrogen column density of log $N$(H\,{\sc I})~$= 21.6 \pm 0.10$~cm$^{-2}$. A more detailed analysis of the optical spectra appears in \cite{Fynbosample}.

\begin{table*}
\begin{tabular}{lllllll}
T$_{\rm start}$ (UT) & T$_{\rm mid}$ (days since T$_{\rm 0,BAT}$) & T$_{\rm exp}$ (s) & filter & magnitude & flux ($\mu$Jy) & telescope + instrument \\ \hline
2008-07-21 10:27:14     & 0.0019 &98.19  &$white$& $14.58 \pm 0.01$&$2945\pm      9.000$& {\it Swift} UVOT \\
 2008-07-21 10:37:15   & 0.0084 & 9.61  &$white$&  $16.73\pm  0.07$&$428\pm       28.0$& {\it Swift} UVOT \\
 2008-07-21 10:39:34   & 0.0105 &98.20   &$white$& $16.95\pm  0.03$&$339\pm      8.00$& {\it Swift} UVOT \\
2008-07-21 11:51:29   & 0.0610 &196.62 & $white$&  $19.12\pm  0.11$&$44.0\pm      4.00$& {\it Swift} UVOT \\
 2008-07-21 12:15:25  & 0.0777 &196.61 & $white$&  $19.40\pm  0.13$&$35.0\pm      4.00$& {\it Swift} UVOT \\
 2008-07-22 04:13:23  &0.9049  &2414.0 & $white$&  $22.39 \pm 0.17$&$6.00\pm      1.00$& {\it Swift} UVOT \\
 2008-07-24 00:50:14 & 2.0790 &1860.5  & $white$& $23.57\pm  0.62 $&$3.00\pm      1.00$& {\it Swift} UVOT \\ \hline
2008-07-21 10:29:01   & 0.0049 &  393.4 &$v$& $ 14.93 \pm 0.01$  &$  3876\pm       53.00$& {\it Swift} UVOT \\
2008-07-21 10:37:54   & 0.0089 &19.4   &$v$& $16.04 \pm 0.13 $   &$  1404\pm       167.0$& {\it Swift} UVOT \\
2008-07-21 10:41:20   & 0.0135 &393.5   &$v$& $16.41 \pm 0.04 $  &$   993\pm       34.0$& {\it Swift} UVOT \\
2008-07-21 11:58:19  & 0.0658 &196.6   &$v$& $18.37\pm  0.24 $   &$  164\pm       36.0$& {\it Swift} UVOT \\
2008-07-21 12:22:16  & 0.0820 &135.6   &$v$& $18.13\pm  0.23 $   &$   204\pm       44.0$& {\it Swift} UVOT \\
2008-07-21 15:23:23 & 0.2118 &797.2   &$v$& $19.93 \pm 0.47 $    &$   39.0\pm       17.0$& {\it Swift} UVOT \\
2008-07-22 04:20:38  & 0.9074 &2524.4  &$v$ &$20.90\pm  0.29 $   &$ 21.0\pm      9.00$& {\it Swift} UVOT \\ \hline
2008-08-01 00:30:33  &    10.5880  &    180  &   $V$   &  25.57  $\pm$  0.23  &  $0.290 \pm 0.061$ &  VLT Antu + FORS2   \\
2008-08-01 00:34:06  &    10.5905  &    180  &   $V$   &  25.64  $\pm$  0.25  &  $0.272\pm 0.063$ &  VLT Antu + FORS2   \\ \hline
2008-07-21 20:51:00  &     0.4444  &    980  &   $R$   &  20.12  $\pm$  0.12  &  $33.3 \pm 3.70$ &  NOT + ALFOSC  \\
2008-07-22 02:36:32  &     0.6747  &     30  &   $R$   &  20.82  $\pm$  0.08  &  $17.5  \pm 1.30$ &  VLT Kueyen + FORS1   \\
2008-07-22 02:41:25  &     0.6781  &     30  &   $R$   &  20.80  $\pm$  0.12  &  $17.8 \pm 2.00$ &  VLT Kueyen + FORS1   \\
2008-07-30 20:57:27  &     9.4533  &   2100  &   $R$   &  25.22  $\pm$  0.50  &  $0.303 \pm 0.140 $ &  WHT + API     \\
2008-08-01 00:37:49  &    10.5931  &    180  &   $R$   &  24.94  $\pm$  0.16  &  $0.393 \pm 0.058 $ &  VLT Antu + FORS2   \\
2008-08-01 00:41:23  &    10.5956  &    180  &   $R$   &  25.01  $\pm$  0.17  &  $0.368 \pm 0.058 $ &  VLT Antu + FORS2   \\
2008-08-10 23:11:45  &    20.5467  &   1080  &   $R$   &  25.75  $\pm$  0.44  &  $0.186 \pm 0.075 $ &  VLT Antu + FORS2   \\
2008-08-20 23:40:32  &    30.5695  &   2700  &   $R$   &  26.47  $\pm$  0.28  &  $0.0960 \pm 0.0250 $ &  VLT Antu + FORS2   \\ \hline
2008-07-21 21:10:12  &     0.4585  &   1800  &   $r$   &  20.43  $\pm$  0.10  &  $31.7 \pm 2.90$ &  LT + RATCAM  \\
2008-07-23 22:40:16  &     2.5211  &   1800  &   $r$   &  22.86  $\pm$  0.13  &  $3.38 \pm 0.400$ &  LT + RATCAM  \\
2008-07-24 21:56:17  &     3.4959  &   1500  &   $r$   &  23.47  $\pm$  0.12  &  $1.92 \pm 0.210 $ &  INT + WFC    \\
2008-07-27 21:30:27  &     6.4759  &    980  &   $r$   &  24.56  $\pm$  0.28  &  $0.705\pm 0.180 $ &  INT + WFC    \\ \hline
2008-08-01 00:45:06  &    10.5981  &    180  &   $I$   &  24.16  $\pm$  0.54  &  $0.593 \pm 0.290 $ &  VLT Antu + FORS2  \\ \hline
2008-07-21 22:17:07  &     0.5050  &   1800  &   $i$   &  20.43  $\pm$  0.22  &  $29.7 \pm 6.00$ &  LT + RATCAM  \\ \hline
\end{tabular}
\caption{Overview of optical photometric observations obtained for GRB\,080721. The magnitudes are $AB$ magnitudes for the $r$ and $i$ filters. The listed magnitudes are not corrected for Galactic extinction; the listed fluxes are corrected for Galactic extinction. Errors are given at the 1$\sigma$ level.}
\label{optdata}
\end{table*}

\section{Measuring the GRB fluence}
\begin{figure}
\begin{center}
\includegraphics[width=6cm, angle=-90]{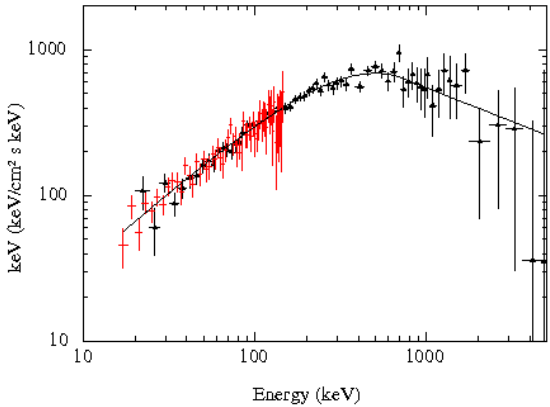}
\caption{
Spectral energy distribution (equivalent to a $\nu F_{\nu}$ representation) showing the 20 keV--7 MeV {\it Konus-Wind} spectrum (black) and the 15--130 keV BAT spectrum (red, scaled by the BAT-{\it Konus-Wind} offset) in the time interval T$_{\rm 0,BAT}$-16~s to T$_{\rm 0,BAT}$+7.81~s. The solid line shows the best-fitting Band function, from which the spectral peak energy can be clearly measured at a few hundred keV.}
\label{batkonussed}
\end{center}
\end{figure}
We performed a joint fit to the 20~keV--7~MeV {\it Konus-Wind} and 15--150~keV BAT spectra, time integrated over the 23.81~s of simultaneous observations (covering the main peaks of gamma-ray emission, Fig. \ref{batkonuslc}). Adopting the Band function model (\citealt{Band}, in which $\alpha$ and $\beta$ are spectral slopes as opposed to the definitions given in Section 2 used elsewhere in this paper) and allowing for a constant normalisation offset between the BAT and {\it Konus} instruments, we find the data are well fitted with $\chi^2$/dof = 105.6/138. We measure a low energy spectral slope $\alpha = -0.96^{+0.08}_{-0.07}$, high energy spectral slope $\beta = -2.42^{+0.22}_{-0.38}$ and spectral peak energy $E_{\rm pk} = 497^{+63}_{-61}$ keV. The $\gamma$-ray spectral energy distribution, showing the location of the spectral peak in the {\it Konus-Wind} energy range at this time, is plotted in Figure \ref{batkonussed}.  
The ratio of BAT to {\it Konus-Wind} normalisation was $0.83\pm0.04$,
with all other model parameters tied. We also split these observations into four time intervals over which we searched for spectral variability. There is no evidence for highly statistically significant ($>3\sigma$) spectral variations within the measured errors, therefore we use the time-averaged spectral fit parameters in further analysis.
We measure a fluence (or time-integrated flux), $S$, of (8.81$^{+0.77}_{-0.75}$) $\times 10^{-5}$ erg cm$^{-2}$ (20~keV--7~MeV) from the joint spectral fit.

\section{Characterising the afterglow}
\subsection{Spectral fits}
The X-ray hardness ratio (HR), defined as the count rate ratio in the energy bands 1.5-10~keV/0.3-1.5~keV, indicates that there are no significant X-ray spectral changes
throughout the entire duration of {\it Swift} XRT observations. We have fit
the HR with a constant value and find a value of HR = 0.78$\pm$0.01 with $\chi^2$/dof = 222.36/246. To allow for an increase or decrease in HR we fit the data
with a linear function and found that this extra degree of freedom does
not improve the fit (which then goes to $\chi^2$/dof = 221.9/245) but is an equally good fit to the constant value assumption.
\begin{figure}
\begin{center}
\includegraphics[width=5.5cm, angle=90]{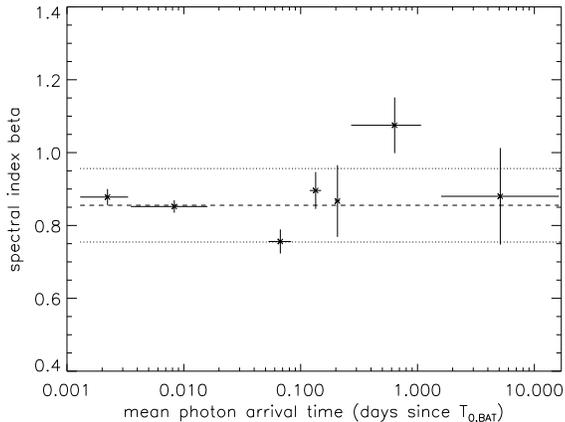}
\caption{The X-ray power law spectral index, $\beta_X$, as a function of mean photon arrival time of the time-resolved spectra. The best-fitting constant value is overlaid (dashed line) together with the standard error between the fit and the data (dotted lines).}
\label{HR}
\end{center}
\end{figure}

To investigate spectral evolution further we created seven time-sliced {\it Swift} XRT spectra covering almost the entire decay:
T$_{\rm 0,BAT}$+112--288~s (WT), 304--1376~s (WT), 4.5--7.2~ks (PC orbit 2), 10.3--12.9~ks (PC orbit 3), 16.9--18.7~ks (PC orbit 4), 23.3--92.6~ks (PC from orbit 5 to 1.16~d) and 137.7--1400~ks (PC from 1.16~d onwards). We fit these spectra simultaneously with an absorbed power law; the Galactic absorption was fixed to $N_{\rm H,gal} = 7\times$10$^{20}$ cm$^{-2}$ \citep{Kalberla}, the intrinsic absorption at $z=2.519$ was tied i.e. was required to have the same value at each epoch, and the power law photon index, $\Gamma = \beta+1$, and normalisation were left as free parameters per epoch. The best fit has $\chi^2$/dof = 1765/1307 ($\chi^2_{\rm red}$ = 1.35). Intrinsic X-ray absorption at the host galaxy is best fitted with $N_{\rm H} = (6.5\pm0.5)\times 10^{21}$ cm$^{-2}$, assuming Solar metallicity, and we note that this is comparable to the column density derived in the optical from hydrogen Lyman-$\alpha$. Plotting the resulting $\beta_X$ values against the mean photon arrival time per spectrum (Figure \ref{HR}) we find that an increasing or decreasing spectral slope is not a significant improvement over a constant value fit. The best fitting constant value is $\beta_X=0.86\pm0.01$, while the standard error between the fit and the datapoints of 0.1 gives a measure of the scatter among the slopes about the constant fit value.

We derived the count rate to flux conversion of 1~count~s$^{-1}$~=~4.1$\times$10$^{-11}$ erg cm$^{-2}$ s$^{-1}$ from the time-averaged spectrum of the XRT PC mode data from 0.5~d onwards (to coincide with the time range covered by our final light curve fits described in the following section).

\begin{figure}
\begin{center}
\includegraphics[width=4.5cm, angle=-90]{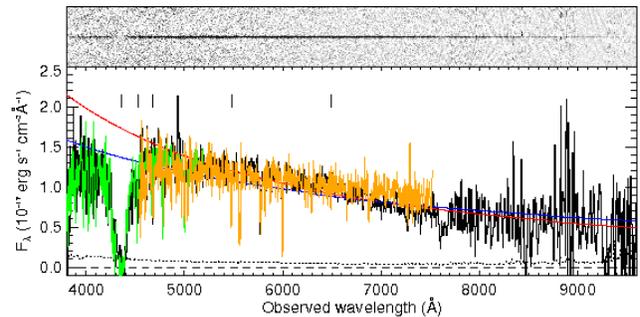}
\caption{Optical spectra taken with the VLT, 0.7~d after the GRB event. The top panel shows the skyline-subtracted 2D spectrum from the 300V grism. The main panel shows the 300V (black), 1200B (green) and 600V (orange) 1D flux calibrated spectra and the 300V error spectrum (dotted line). The H\,{\sc I} Ly$\alpha$ and Si\,{\sc II} lines used in metallicity determination are indicated above. The solid lines overplotted show power law fits for the cases of $\beta_O=\beta_X$ (blue) and $\beta_O=\beta_X-0.5$ (red), not including any host galaxy extinction.}
\label{optspec}
\end{center}
\end{figure}
An estimate for the spectral index between optical $R$ and X-rays (centred at 1.7 keV) after 0.5~d results in $\beta_{OX} \approx 0.65$. Note that this estimate does not take into account any optical extinction in the host: we feel that our late-time optical data suffer from too much noise
to reliably estimate any extinction in the host galaxy, while the assumption of a standard fireball model may not be valid at the time of the (early) UVOT data where extinction may be best measured.  
We also attempt to estimate the optical spectral index $\beta_O$ using our optical spectra taken 0.7~d after the GRB event. The three spectra taken with the VLT are shown in Fig. \ref{optspec}. We overlay on these a power law with spectral slope equal to that found in the X-rays ($\beta=0.86$) and a power law with slope shallower by 0.5 ($\beta=0.36$) and see that both may be accommodated when all three spectra are included. Again, intrinsic extinction is not included in these power law models and would be impossible to disentangle from the underlying spectral slope in this case. Thus we conclude we cannot reliably estimate $\beta_O$. 

\subsection{Light curve fits}
\begin{figure}
\begin{center}
\includegraphics[width=8.5cm, angle=0]{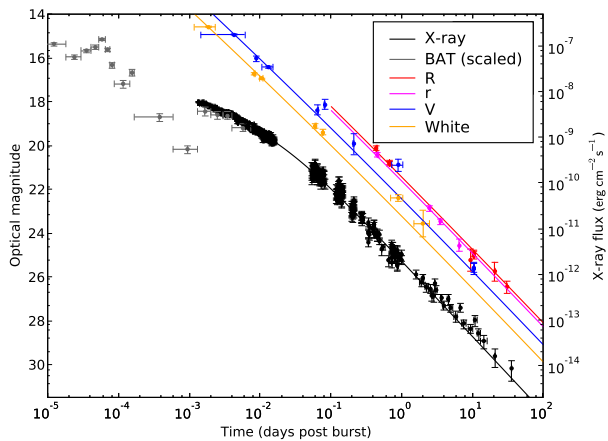}
\caption{$\gamma$-ray (grey, BAT only), X-ray (black, final 2 points from {\it XMM-Newton}) and optical (colours: orange = $white$, blue = $V$ and $v$, red/pink = $R/r$) light curves shown together. The solid lines show a smoothly broken power law decay fitted to all bands from 430~s onwards (see Table \ref{table:lcfit}). The optical magnitudes are corrected for Galactic extinction, while the X-ray fluxes are 0.3--10 keV observed fluxes. We have transformed the $\gamma$-ray fluxes to the X-ray band using an absorbed power law extrapolation.}
\label{xandoptlc}
\end{center}
\end{figure}

To obtain an estimate of some of the blastwave parameters, we fit the light curves and compare the results with the spectral slope for the X-rays. The full optical and X-ray light curves are shown in Fig. \ref{xandoptlc}. We can obtain a good fit to the X-ray light curve using a smoothly broken power law, as suggested by e.g. Beuermann et al. (1999; subsequently explored further in \citealt{Granot}), which allows for a smooth transition from one power law to another. We use the formulation 
\begin{equation}
f(t) \propto  t^{\alpha_2} \times { \left ( \frac{ 1 + {( t / t_{\mathrm{b}}
)}^{(s \cdot \gamma )} }{ 1 +  {( 1 / t_{\mathrm{b}} )}^{(s \cdot \gamma)}
} \right ) } ^ {( -1 / s)},
\end{equation}
with $\alpha_1$ and $\alpha_2$ the decay parameters before and after the
break respectively, $\gamma = \alpha_2 - \alpha_1$, $t_{\mathrm{b}}$ the
break time and $s$ a ``smoothness'' parameter for the break (fixed at the value of 1 throughout our fits).
We omit the first 430~s of XRT data from the fitting, where we see deviations from a power law shape (e.g. \citealt{Marshall}). Overlap between the XRT and BAT light curves at these very early times also suggests the prompt emission may contribute significantly. Thereafter, the light curve decays in a very smooth fashion. 

We simultaneously fit the optical data, where we keep the decay parameters for each optical band tied together. 
We initially employed models where the decay indices and break time of the X-ray light curve are free parameters, and the optical decays either as a smoothly broken power law as well (with the break time the same as that in X-rays, Fig. \ref{xandoptlc}), or a single power law. The resulting difference between the latter two scenarios is minimal, but the preferred scenario is the smoothly broken power law for both the X-ray and optical light curves, which improved the fits over a sharply broken power law fitted to the optical light curve (the F-test gives a probability of this result being obtained by chance of 0.4\%). Based on these initial results, we see a difference at late times between optical and X-ray decay slopes of $\Delta\alpha\sim0.25$, which is expected in one particular blastwave scenario (see the following section). We then fitted the models again, but this time we constrained the late-time light curves with a fixed 0.25 slope difference between the X-ray and the optical regime. The resultant fit parameters are shown in Table \ref{table:lcfit}. 

\begin{table*}
\begin{tabular}{p{2.8cm}|llllll}
model & $\alpha_{X,1}$ & $\alpha_{X,2}$ & $\alpha_{O,1}$ & $\alpha_{O,2}$ &
$t_{\mathrm{break}}$ (d) & $\chi^2$/dof \\
\hline
data from 430~s: & & & & & & \\
sbpl$_{X}$, pl$_{O}$ & $0.630 \plusminus{0.046}{0.049}$ & $1.656
\plusminus{0.026}{0.025}$ & $1.272 \pm 0.010$ & --- & $0.0210
\plusminus{0.0039}{0.0032}$ & $508.4/411$ \\
sbpl$_{X}$, sbpl$_{O}$ & $0.630 \pm 0.029$ & $1.656 \pm 0.015$ &
$1.174\plusminus{0.031}{0.025}$ & $1.369\plusminus{0.027}{0.031}$ &
$0.0210\plusminus{0.0039}{0.0031}$ & $504.3/410$ \\
sbpl$_{X}$, pl$_{O}$ with $\alpha_{X,2} = \alpha_{O} + 0.25$ &
$0.564\plusminus{0.002}{0.010}$ & $1.616 \plusminus{0.078}{0.164}$ &
$1.366{}^1$ & --- & $0.0157 \pm 0.0013$ & $687.0/412$ \\
sbpl$_{X}$, sbpl$_{O}$ with $\alpha_{X,2} = \alpha_{O,2} + 0.25$  & $0.630
\plusminus{0.005}{0.006}$ & $1.656 \plusminus{0.019}{0.006}$ &
$1.153\plusminus{0.019}{0.018}$ & $1.406{}^1$ &
$0.0210\plusminus{0.0023}{0.0029}$ & $506.2/411$ \\ \hline
data from 0.5~d: & & & & & & \\
pl& ---& $1.428\plusminus{0.035}{0.034}$ & --- &
$1.410\plusminus{0.041}{0.039}$ & ---& $55.3/58$ \\
pl, with $\alpha_{X,2}=\alpha_{O,2}$ & --- & $1.428\plusminus{0.022}{0.035}$
& --- & $1.428{}^1$ & ---& $55.6/59$ \\
pl, with $\alpha_{X,2}=\alpha_{O,2}$+0.25 & --- &
$1.428\plusminus{0.251}{0.011}$ & --- & $1.178{}^1$ &---& $98.7/59$ \\
\hline
\end{tabular}
\caption{Results of fits to the X-ray and optical light curves. sbpl = smoothly broken power law, pl = power law. ${}^1$Constrained parameter, so the errors are taken from $\alpha_{X,2}$.}
\label{table:lcfit}
\end{table*}

We then concentrate our efforts on estimation of the light curve decay of the very late-time data only, where we can be more confident there is little to no influence of possible other components, including flares and energy injection. From 0.5~d onwards, we measure an X-ray decay of $\alpha_{X,2}\sim 1.4$ (Table \ref{table:lcfit}). This is interestingly and significantly different from the $\alpha_{X,2} \sim 1.6$ found when we fit the whole dataset from 430~s onwards (in which a single break is best fitted at 0.02~d), and indicates a change in slope which is not obvious when combined with earlier data. 
Fitting the optical and X-ray decay indices independently rather than tied shows a difference of only $0.03$, while any difference owing to a so-called cooling break (see the following section) between optical and X-rays should be $\Delta\alpha = 0.25$ in the fireball model \citep[e.g.][]{Sari}. We thus conclude there is no such spectral break in this scenario. 

The question arises as to which of the two fits is actually better: the one
containing data from 430~s onwards (smoothly broken power law), or the one
with data only after 0.5~d (single power law). For this, we calculate the final $\chi^2$ value, only for the data
past 0.5~d, for both fits. That is, the fits themselves are still done
by minimizing $\chi^2$ calculated from the full applicable data range (post
430 s and post 0.5~d, respectively),
but we compare $\chi^2$ calculated by using only data past 0.5~d for
each fit (where likely only a power law applies). The $\chi^2$/dof for the
smoothly broken power law in this case is rather bad, 111.1/55,
compared to 55.3/58 for a power law only. This would indicate that the
relatively good total $\chi^2$ value for the smoothly broken power law
fit arises largely from the early data between 430~s and
0.5~d. This could be the case if errors in the early section are
slightly overestimated. It could also indicate a second break at
X-ray wavelengths in this interval, which would account for the difference in
final power law decays ($\alpha_{X}$ of 1.6 versus 1.4). Such a break would, however,
increase the complexity of the underlying light curve model, and may
make it more difficult to interpret the results. Lastly, we could also
be seeing effects from fitting a smoothly broken power law, where the
second decay index is the actual final decay index, and thus may not
be equal to the decay index measured along a small portion of the
light curve \citep[see e.g.][]{johannesson2006}.

\section{Consistency with the fireball model}

We now investigate whether or not the data are consistent with the fireball model predictions.
We use the closure relations (found in, e.g., Table 1 of \citealt{Zhang}) relating the optical and X-ray spectral and temporal slopes to determine how this GRB afterglow may fit into the standard fireball model. 
In Section 4.2 we have performed the same light curve fits to two subsets of the data: one beginning at 430~s since the GRB and one beginning 0.5~d since the GRB, with two different outcomes. Following the reasoning in the last paragraph of Section 4.2, we examine the results obtained from fitting the data from 0.5~d to a single power law.

The late-time temporal fits show that the slopes are essentially the same in the X-ray and optical bands, with $\alpha \approx 1.40$, which indicates that both observing bands are in the same spectral regime, either both below or above the cooling break. In the case where they are both below the cooling break and the circumburst medium is homogeneous, the value of $p$ can be derived to be $2.72 \pm 0.20$ from $\beta_X$ (well within the range of previously found values, \citealt{Shen,Starling}), and the predicted temporal slope is $\alpha = 1.29 \pm 0.15$, which is consistent with the observed value. A stellar-wind-like medium or a scenario in which both the optical and X-ray bands are situated above the cooling frequency can be ruled out by the standard closure relations between the spectral and temporal indices.
The value for the X-ray spectral index, however, is somewhat at odds with the estimated $\beta_{OX}$, if there is no supposed break across the $\beta_{OX}$ range. Host galaxy extinction may mean the intrinsic optical magnitudes are brighter than those observed, allowing an increase in $\beta_{OX}$, which could resolve this apparent discrepancy.
To investigate this, we can combine the $VRI$ optical photometry (Table \ref{optdata}) with an X-ray spectrum to estimate extinction from the broadband spectral energy distribution (SED) in the case that $\beta_O = \beta_X$. The SED, centred at 1.725~d, was created and fitted via the method outlined in \cite{Starlingcolumns} and adopts a single absorbed power law as the underlying model and Small Magellanic Cloud-like host galaxy extinction. We find that extinction is required at the level of $E(B-V)=0.21\pm0.07$, corresponding to $A_V \sim 0.6$ magnitudes. This value is one of the highest measured for a GRB host galaxy \citep{Starlingcolumns,Schadycolumns}, but such measurements are made only for GRBs with detected optical afterglows and presumably extinction is much higher in the optically undetected dark bursts \cite[e.g.][]{Cenko,Fynbosample}.

For completeness we performed the same tests on the best-fitting smoothly broken power law model to the data from 430~s onwards. We find that the power law index of the electron energy distribution, $p$, derived from the light curve predicts harder spectral indices than are observed. We investigated the possibility of an additional spectral component, caused by Inverse Compton (IC) emission, which would harden the spectrum (as also proposed for e.g. GRB\,990123, \citealt{Corsi} and GRB\,000926, \citealt{Harrison}). Following the equations in \cite{SariEsin} and \cite{Corsi}, we find that the closure relations for the temporal and spectral indices still can not be satisfied.
We note that in this scenario the break at 0.02 d may be considered an early jet break. There are two arguments against attribution of this feature to the jet break. Firstly, after a jet break the temporal slopes are given by $\alpha = p$,
regardless of whether the observing frequency is above or
below the cooling frequency, so the optical and X-ray temporal
slopes should be the same. This is not the case for this dataset. Secondly, if this feature were a jet break it would imply an opening angle of 0.44$^{\circ}$ (calculated using the equations in the following section), far smaller than most previously determined values \citep[e.g.][but see also \citealt{Schady,Racusin}]{Frail,Berger}.

Concluding, the results from fits to the late-time only data indicate that both observing bands are in the same spectral regime and in order to correctly predict the X-ray spectral slope the jet must be traversing a homogeneous medium. The difference between the optical-to-X-ray spectral index and the X-ray spectral slope is not predicted, but could be explained by the invocation of host galaxy extinction of $A_V \approx 0.6$. The results from using the full range of data can not be satisfactorily described by the fireball model and/or with an early jet break, and it is likely that in this case a single smoothly broken power law model is not applicable.  

\section{Energetics}
These data have shown that a late-time temporal break at X-ray or optical wavelengths is not required. We can therefore use the latest observation time as a lower limit to any jet break time, and calculate the required limits on opening angle and total energy budget for the GRB jet.
To calculate the isotropic energy $E_{\rm iso}$ we use the following equation
\begin{equation}
E_{\rm iso}(\gamma) = \frac{4\pi\cdot S\cdot d_l^2}{1+z}, 
\end{equation}
adopting the values listed in Sections 2 and 3 ($z=2.591 \pm 0.001$, fluence $S= 8.81^{+0.77}_{-0.75} \times 10^{-5}$ erg cm$^{-2}$) and a luminosity
distance $d_l$ of 6.5$\times$10$^{28}$ cm (where $H_0 = 71$ km s$^{-1}$ Mpc$^{-1}$, $\Omega_{\rm M} = 0.3$ and $\Omega_{\Lambda} = 0.7$, adopted in order to compare our value with the sample of \citealt{KocevskiButler}).
We find $E_{\rm iso}=1.30^{+0.12}_{-0.11} \times 10^{54}$ erg. We have not included a $K$~correction, because the energy range used encompasses the peak energy,
and the measured fluence should be sufficiently close to the bolometric
fluence\footnote{For comparison, the restframe 1--10000~keV fluence is 7.99$^{+0.35}_{-0.36} \times 10^{-5}$ erg cm$^{-2}$ corresponding to $E_{\rm iso,rest} = 1.18\pm0.05 \times 10^{54}$ erg}.

Using this value for $E_{\rm iso}$ we can go on to calculate an upper limit on
the jet opening angle \citep[e.g.][]{Frail}, since no jet break is seen up to our last {\it XMM-Newton} observation. This is assuming we may apply the fireball model for the case of a uniform jet (see Section 5 for further discussion of this).

\begin{eqnarray}
\label{thetaj}
\theta_{j} = 0.057\left(\frac{t_j}{1 {\rm d}}\right)^{3/8} \left(\frac{1+z}{2}\right)^{-3/8}\left (\frac{E_{\rm iso}}{10^{53} {\rm erg}}\right)^{-1/8} \nonumber\\
 \cdot\left(\frac{\eta_{\gamma}}{0.2}\right)^{1/8}\left(\frac{n}{0.1 {\rm cm}^{-3}}\right)^{1/8}.
\end{eqnarray}
Adopting typical values for efficiency and density of $\eta_{\gamma}=0.2$ and $n=0.1$~cm$^{-3}$, we find $\theta_j \ge
0.127^{+0.002}_{-0.001}$ radians or ($7.30\pm0.08)^{\circ}$.\\ %0.1274, range = 0.126-0.129 

The lower limit on the total energy budget in $\gamma$-rays for the GRB is then obtained from the following
equation which accounts for jet collimation using the newly derived jet opening angle upper limit,
\begin{equation}
E_{\gamma} = E_{\rm iso}(1 - \cos \theta_j),
\end{equation}
which gives $E_{\gamma}  \ge (1.06^{+0.07}_{-0.07}) \times 10^{52}$ erg, or more conservatively taking the lower limit for $E_{\rm iso}$ then $E_{\gamma} \ge 9.88 \times 10^{51}$ erg using the end of the final {\it XMM-Newton} observation (36.08 d or 3.12$\times$10$^{6}$~s) as our lower limit on any jet break time.

\section{Discussion}
\subsection{Summary of results and associated caveats}
We have followed the afterglow of the high $E_{\rm iso}$, relatively high redshift GRB\,080721 to beyond 10$^{6}$~s at both X-ray and optical wavelengths. The late-time decay is well fitted with a single power law with no requirement for a break. This suggests that the jet break must lie beyond the detection limits of the intruments used, and from this we calculate a lower limit to the total prompt $\gamma$-ray energy of the GRB of $E_{\gamma} \ge 9.88 \times 10^{51}$ erg and to the jet opening angle of $\theta_j \ge 7.22^{\circ}$. This relies on the assumption that the fireball model describes the physics of the jet, the jet is uniform in structure, and that the standard values for surrounding density and efficiency of energy conversion are applicable. We note that the dependence of the total energy on these latter two parameters is very weak (Equation 2). We have compared the closure relations from the fireball model with the observed temporal and spectral slopes. The model and the observations cannot be reconciled for the whole dataset, with departures from the model at early times (before half a day), and different temporal slopes are derived depending on whether data before 0.5~d are included in the fit. It is likely that additional components combine with the afterglow emission at the onset of the afterglow, rendering the fireball model inapplicable, but at what time the afterglow begins to dominate is not known. This illustrates the difficulties of pinning down the details of the blastwave physics even when broad-band well-sampled data are available.
The data either from 0.5~d or from 430~s onwards require that the GRB jets traverse a homogeneous circumburst medium. The apparent inconsistency between the X-ray spectral index and the optical to X-ray spectral index may be caused by extinction in the host galaxy.  

We have also investigated the alternative scenario in which an earlier break in the light curve at $\sim$0.02~d is the jet break, and found that this can not work within the fireball model. If we assume the fireball model, and therefore that one of the measured observational parameters is contaminated (perhaps by an additional component or incorrect other assumption), the implied jet opening angle would be very small (0.44$^{\circ}$) for a GRB (or indeed any known jet source). Such collimation cannot be explained by any current models and the chances of the narrow jet happening upon our line-of-sight are small. We do not prefer this scenario for these reasons, but we note that this explanation has been the preferred model for some other GRBs which we discuss in Section 7.3.
In the following subsection we summarise the main classes of long GRB progenitor models, and discuss whether any of these can reach energies 10$^{52}$ erg and beyond.

\subsection{Implications for GRB progenitor models}
In the collapsar scenario, two different ways to extract energy to make a GRB are suggested
in the literature, one using neutrinos and the other harnessing magnetic fields.

In the neutrino-driven model, the energy that can be extracted depends sensitively on the accretion rate of collapsing material onto the newly formed black
hole through the accretion disc and on the Kerr parameter describing the rotation of the black hole, as seen in Table 3 of Popham et al. (1999; see also \citealt{MacFadyen,Aloy,Nagataki}). For typical long GRB durations (tens to hundreds of seconds) a GRB total energy of 10$^{52}$~erg is certainly possible within this model particularly for fast rotation rates. The observational testing of such models depends, therefore, not on measurements of GRB total energies but on determination of the rotation and mass accretion rates. According to the recent GRB progenitor models by \cite{Yoon}, \cite{YoonLangerNorman} and \cite{Woosley}, a lower metallicity star is preferred for a higher Kerr parameter.
\cite{Yoon} and \cite{YoonLangerNorman} introduced models of chemically homogeneous evolution which produces helium stars with little or no hydrogen envelope that are metal-poor fast-rotators, and can therefore go on to form a GRB. The complexities of stellar evolution are not all well understood, and make the determination of GRB progenitors a difficult undertaking. 

However, the identification of the crucial role of metallicity opens up an avenue for observational constraints. We can measure the metallicity of the circumburst medium or host galaxy along the line-of-sight to GRBs through afterglow spectroscopy, and for most long GRBs metallicities of order a tenth the Solar value have been inferred from this method \citep{Fynbo,Prochaska}, with the lowest measured value being $\frac{1}{100}$ Solar for GRB\,050730 \citep{Chen,Starling050730,Prochaska} though lines are often saturated hence likely to provide only lower limits on the metallicity \citep{Prochaska06}. We can measure the metallicity, $Z$, of the environment of GRB\,080721 from the column densities of neutral hydrogen (H\,{\sc I} Ly$\alpha$) and singly ionised silicon (Si\,{\sc II} $\lambda$1808 \AA) in absorption (Fig. \ref{optspec}) of [Si/H]$>$-1.1. All the other observed spectral lines are likely to be saturated, hence we have used the weakest detected metal line. This equates to $Z\sim0.08$ times the Solar value, within the typical distribution for GRBs. While this should strictly be treated as a lower
limit due to possible saturation of the Si\,{\sc II} $\lambda$1808 \AA\ line, saturation is likely to be moderate at most and, therefore, the actual metallicity of the absorber should be of the order of one tenth of solar, unless significant dust depletion effects are at play (as silicon is a mildly refractory element).
If the metallicity is significantly lower than 0.1 Solar, the Kerr parameter may be close to $a=1$ \citep{Yoon,YoonLangerNorman}, and an energy of 1$\times$10$^{52}$ erg could be easily achieved according to the calculations of \cite{Popham}.
A high metallicity could be an indication of large initial stellar masses, or simply that the models are not applicable. 

The magnetic-field-driven model uses the Blandford-Znajek mechanism to extract rotation energy from the black hole \citep{BlandfordZnajek}. This idea is indeed supported by the recent general relativistic magnetohydrodynamical simulations by \cite{Barkov} and \cite{Komissarov}.
The energies that can be extracted are up to 10$^{54}$ erg at 100\% efficiency. Typical efficiencies are likely to be far lower than this \citep[but see][]{Komissarov}, and with 1\% efficiency the lower limit on total energy of GRB\,080721 can be accommodated. Here again the Kerr parameter is an important factor in determining the available energy, in which case lower metallicity progenitor stars may be preferable.

Interestingly, the measured energy limit for GRB\,080721 is comparable to the energy budget in the magnetar scenario \citep{Thompson}, where the extraction of the rotational energy of a millisecond pulsar by strong magnetic fields is suggested \citep{Duncan,Wheeler,Thompson,KomissarovBarkov}. The magnetar scenario for long GRBs has received renewed interest in recent observational works, where it has been suggested to explain some plateau phases and very steep X-ray declines \citep[][specifically in e.g. GRB\,070110, \citealt{Troja,Lyons}]{ZhangMeszaros,Rosswog}. The claim of an 8~s periodicity in the highest energy bands for GRB\,090709A suggests association with a magnetar for this source \citep{Golenetskii09,Gotz,Markwardt,Ohno}.
The theoretical prediction for the upper limit on the energy of a maximally rotating neutron star is few$\times$10$^{52}$ erg \citep{Thompson}, and we note that $\sim$10$^{52}$ erg is the lower limit we derive for this GRB implying very high efficiency close to 100\%. We therefore conclude that it is unlikely that GRB\,080721 is produced via magnetar formation.

\subsection{Comparison with other long GRBs}
\begin{figure}
\begin{center}
\includegraphics[width=4.5cm, angle=90]{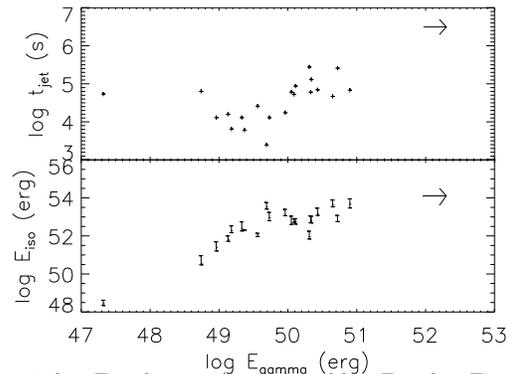}
\caption{log $E_{\gamma}$ -- log $t_{\rm jet}$ (upper) and log $E_{\gamma}$ -- log $E_{\rm iso}$ (lower) plots for {\it Swift} long GRBs with jet breaks or possible jet breaks (taken from Table 2 of \citealt{KocevskiButler}; redshifts may be spectroscopic, photometric or from host association). Jet break time is given in the observer frame. The arrows in the upper right of each plot show the limits set here on GRB\,080721.}
\label{EgammaEisoSwift}
\end{center}
\end{figure}

The implied total energy for GRB\,080721 based on the lower limit calculated here is not only a challenge to some progenitor models, but is a rare occurrence among GRBs. We have compared this limit to the energies measured from detections of jet breaks in a {\it Swift} sample of GRBs published by \cite{KocevskiButler} in Fig. \ref{EgammaEisoSwift}. The lower limit for GRB\,080721 is much larger than any energy measured from a jet break in this sample. Comparing this source with the Amati relation \citep{Amati} we see that it falls at the high $E_{\rm peak}$--$E_{\rm iso}$ end of the correlation, in a continuation of the long GRB population out to the most energetic sources. The {\it XMM-Newton} observations have enabled a longer baseline for jet searches and this source was chosen specifically for its unusually high isotropic energy.

GRB\,061007 \citep{Mundell,Schady} was also a highly energetic GRB with an isotropic energy of 1$\times$10$^{54}$ erg \citep{Golenetskii06}. The closure relations for this source were also not entirely satisfied and no spectral evolution was found from 80~s to 3.1~d after the trigger. The high implied total energy, in both the wind and homogeneous medium cases considered in \cite{Schady} led the authors to conclude that an early jet break must have occurred, before T$_0$+80~s, which implies a highly collimated outflow with opening angle $0.1$--$0.8^{\circ}$, four times smaller than any determined previously. This scenario, which we investigated for GRB\,080721 in Section 5 and subsequently ruled out, requires that the narrow jet happens to lie along our line-of-sight, may be supported by the rarity of such non-evolving afterglows and the unusually high initial optical brightness for GRB\,061007. However, we discussed long GRB progenitor models in the previous section for which the large energies measured both for GRB\,061007 and in this paper for GRB\,080721 when assuming a very late, unobserved jet break with a more typical jet opening angle, are in fact achievable. 

Another case for which the observations implied a high energy budget, while the standard fireball model was a poor description of the afterglow, was GRB\,080319B \citep{Racusin}. This collapsar scenario for this long GRB is, however, supported by the detection of a supernova signature \citep{Tanvir}. To explain the complex afterglow behaviour, not smooth and featureless in this case but appearing to have a number of components including an initially extremely bright optical counterpart, \cite{Racusin} introduced a second jet. This 2-jet model consisted of an inner jet with $t_{\rm jet}$~=~0.03~d and opening angle 0.2$^{\circ}$ and an outer jet more typical of GRBs with $t_{\rm jet}$~=~11.6~d and opening angle 4$^{\circ}$. Double and even triple jet systems are occasionally invoked when single jet models simply cannot explain the observations (e.g. GRB\,030329, Berger et al. 2003; GRB\,021004, Starling et al. 2005a; GRB\,050401, Kamble et al. 2009; GRB\,050802, Oates et al. 2007, De Pasquale et al. 2009), and while it is likely to be more physical to assume some structure across the jet \citep[e.g.][]{Rossi,ZhangMes2002}, the interaction of multiple jets is not well known. 

\section{Conclusions}
We have observed the bright, highly energetic afterglow of GRB\,080721 out to 36 days since the
trigger in the optical and X-ray bands. We conclude that no jet break is present in the late-time light
curve and we rule out a jet-break origin for the early light curve break, inferring a limit on the opening angle of $\theta_j \ge 7.22^{\circ}$ and setting constraints on the total energy
budget of the burst of $E_{\gamma} \ge 9.88\times 10 ^{51}$ erg within the fireball model. To obey the fireball model closure relations the GRB jet must be expanding into a homogeneous medium and be extincted in the optical bands by approximately 0.6 magnitudes. 
The energy constraint we derive can be used as observational input for models of the progenitors of long gamma-ray bursts. We can likely rule out a magnetar progenitor for this GRB as this would require close to 100\% efficiency to reach the observed total energy. Such high collimation-corrected energies could be accommodated with certain parameters of the standard massive star
core-collapse models. One of the key observational parameters in distinguishing between various core-collapse models is the metallicity, diagnostic only if it lies outside of the typical range of $0.01<Z<0.1$. The metallicity we measure for GRB\,080721 unfortunately does not allow a distinction between models to be made.

The occurrence of such highly energetic or narrowly beamed GRBs is rather rare, and for those which have the required measurements to allow a study such as this one on GRB\,080721 we estimate there may be a handful per year among the 100--130 well localised GRBs currently triggering operational satellites. Measurement of the high energy spectral peak, securing the redshift and monitoring the light curves for as long as possible (preferably in multiple bands) are all crucial, and these data have shown that a full understanding of the blastwave physics can be difficult to achieve even with a good quality, well-sampled, broad-band dataset. To determine whether these sources are true outliers from the GRB population, perhaps requiring different or more extreme progenitors, will require dedicated efforts to build up small samples to compare with each other and with the growing database of more typical GRBs.

\section{Acknowledgments}
We acknowledge useful discussions with Andrew Read, Jenny Carter, Daniele Malesani and Tayyaba Zafar. RLCS, ER, KLP, KW, PAC and JPO acknowledge financial support from STFC. AJvdH is supported by an appointment to the NASA Postdoctoral Program at the MSFC, administered by ORAU through a contract with NASA. S-CY is supported by the DOE SciDAC Program (DOE DE-FC02-06ER41438). PJ acknowledges support by a Marie Curie European Re-integration Grant within the 7th European Community Framework Program under contract number PERG03-GA-2008-226653, and a Grant of Excellence from the Icelandic Research Fund. This work made use of data supplied by the UK {\it Swift} Science Data Centre at the University of Leicester and {\it XMM Newton} data taken under AO-7 Proposal No. 055597. The {\it Konus-Wind} experiment is supported by a Russian Space Agency contract and RFBR grant 09-02-00166a. Based on observations made with ESO Telescopes at the Paranal Observatory under programme ID 081.D-0853. The INT and WHT are operated on the island of La Palma by the Isaac Newton Group in the Spanish Observatorio del Roque de los Muchachos of the Instituto de Astrof\'isica de Canarias. The Liverpool Telescope is operated on the island of La Palma by Liverpool John Moores University in the Spanish Observatorio del Roque de los Muchachos of the Instituto de Astrofisica de Canarias with financial support from the UK Science and Technology Facilities Council. Based on observations made with the Nordic Optical Telescope, operated on the island of La Palma jointly by Denmark, Finland, Iceland, Norway, and Sweden, in the Spanish Observatorio del Roque de los Muchachos of the Instituto de Astrof\'isica de Canarias. We thank Birgitta Nordstrom and Edita Stonkute for performing the NOT observations.

\bsp

\label{lastpage}

\end{document}